\documentclass[aps,pre,onecolumn,groupedaddress]{revtex4}
\usepackage{graphicx} 
\usepackage{dcolumn} 
\usepackage{bm}
\newcommand{\ignore}[1]{} 
\usepackage{epsfig} 
\usepackage{color}
\usepackage{amsmath,amssymb}

\begin{document}

\title{Contrarian Majority rule model with external oscillating propaganda and individual inertias}

\author{M. Cecilia Gimenez}
\affiliation{IFEG (CONICET), FaMAF (UNC), C\'ordoba, Argentina; ceciliagim@gmail.com}

\author{Luis Reinaudi}
\affiliation{INFIQC (CONICET), Fac. de Ciencias Qu\'imicas (UNC), C\'ordoba, Argentina; luis.reinaudi@unc.edu.ar}

\author{Serge Galam}
\affiliation{CEVIPOF--Centre for Political Research, Sciences Po and
CNRS, 1, Place Saint Thomas d'Aquin, 75007 Paris, France; serge.galam@sciencespo.fr}

\author{Federico Vazquez}
\affiliation{Instituto de C\'alculo, FCEyN, Universidad de Buenos Aires and Conicet, Intendente Guiraldes 2160, Cero + Infinito, Buenos Aires C1428EGA, Argentina; fede.vazmin@gmail}

\date{\today}

\begin{abstract}
We study the Galam majority rule dynamics with contrarian behavior and an oscillating external propaganda, in a population of   agents that can adopt one of two possible opinions.  In an iteration step, a random agent interacts with other three random agents and takes the majority opinion among the agents with probability $p(t)$ (majority behavior) or the opposite opinion with probability $1-p(t)$ (contrarian behavior).  The probability of following the majority rule $p(t)$ varies with the temperature $T$ and is coupled to a time-dependent oscillating field that mimics a mass media propaganda, in a way that agents are more likely to adopt the majority opinion when it is aligned with the sign of the field.  We investigate the dynamics of this model on a complete graph and find various regimes as $T$ is varied.  A transition temperature $T_c$ separates a bimodal oscillatory regime for $T<T_c$ where the population's mean opinion $m$ oscillates around a positive or a negative value, from a unimodal oscillatory regime for $T>T_c$ in which $m$ oscillates around zero.  These regimes are characterized by the distribution of residence times that exhibits a unique peak for a resonance temperature $T^*$, where the response of the system is maximum.  An insight into these results is given by a mean-field approach, which also shows that $T^*$ and $T_c$ are closely related.
\end{abstract}

\maketitle


\section{Introduction}
\label{introduction}

In the last decades, statistical physics has expanded its scope to
venture into the field of sociology, giving rise to a discipline called \textit{sociophysics} 
\cite{Galam-1982,Weidlich,Stauffer,galam-1999,Galam2,Galam3,Galam-2004,Axelrod,Axelrod-2,Vazquez-2022}. 
A commonly studied phenomenon is the dynamics of opinion formation, by means of simple mathematical models.  In these models, individuals are called agents, 
and each of them is characterized by the value of a variable that
represents its opinion on a particular topic --such as the intention
to vote for a candidate in a ballot-- which, for simplicity, can take
one of two possible values ($+1$ or $-1$).  The opinion of each agent
can change after interacting with other agents following simple rules.
One of the most implemented interaction rule is that introduced in a model by
Galam \cite{galam-2002} and extensively studied later on 
\cite{Galam3,majority-Redner,Mobilia,Kuperman-2002}, to which we refer as
the Galam Majority Model (GMM), in which all agents of a group 
chosen at random adopt the opinion of the majority in that group.  This local dynamics drives a steady increase of the initial global majority opinion (provided the system's symmetry is not broken at ties for even size groups) which eventually ends at a consensus, i.e., an absorbing state where all agents share the same opinion.  Multiple extensions of the GMM have been studied in
the literature, including the possibility of a contrarian behavior, that is, all
members of a chosen group taking the minority opinion \cite{Galam-2004}.
This work studied the effects of introducing a fixed fraction $a$ of
contrarian agents on the original GMM, where it was found that,
instead of a frozen consensus as in the model with no contrarians, the system reaches an ordered stationary state for $a<a_c$ and a disordered
stationary state for $a>a_c$.  The transition value $a_c$ separates an
ordered phase where a large majority of agents hold the same opinion,
from a disordered phase in which both opinions are equally
represented in the population.   

Many other opinion formation models with contrarians were also
studied in 
\cite{Stauffer-2004,Schneider-2004,Lama-2005,Lama-2006,Sznajd-2011,Nyczka-2012,Gimenez-2012,Gimenez-2013,Masuda-2013,Banisch-2014,Banisch-2016,Khalil-2019,Martins-2010,Li-2011,Tanabe-2013,Yi-2013,Crokidakis-2014,Guo-2014,Gambaro-2017,Gimenez-2022}.
In particular, the effects of contrarian behavior was also investigated in the voter model (VM)
for opinion formation \cite{Banisch-2014}, where agents interact by
pairs and one adopts the opinion of the other with probability $1-p$
(imitation) or the opposite opinion with probability $p$ (contrarian).
It was shown that the model displays a transition from order to
disorder when the probability of having a contrarian behavior
overcomes the threshold $p_c=(N+1)^{-1}$ in a system of $N$ agents.  The contrarian voter model \cite{Banisch-2014} was recently studied under the presence of a mass media
propaganda that influences agents' decisions \cite{Gimenez-2022}.  The
propaganda was implemented in the form of an external oscillating field
that tends to align agents' opinions in the direction of the field.
It was found a stochastic resonance (SR) phenomena within an oscillatory
regime, that is, there 
is an optimal level of noise for which the population effectively
responds to the modulation induced by the external field 
\cite{Gammaitoni-1998, Gammaitoni-2009}.

In order to expand our knowledge on the combined effects of
contrarians and propaganda on opinion models, we study in this article
the GMM with contrarian behavior under the presence of an external
field.  Each agent in the population can either follow a majority rule
that increases similarity with its neighbors or behave as a
contrarian by adopting the opposite opinion, with respective probabilities $p(t)$ and $1-p(t)$.  The
majority probability $p(t)$ varies in time according to an external
field, based on a mathematical form introduced in
\cite{Gimenez-2012,Gimenez-2013} for the Sznajd model and implemented in
\cite{Gimenez-2021,Gimenez-2022} for the VM, so that agents tend to follow the
majority when it is aligned with the field.  By exploring the dynamics
of the GMM model under the influence of an oscillating external field
and the presence of contrarians, we aim to gain deeper insights into
the manifestation of the SR phenomenon in opinion dynamics models.  We
show that this model exhibits unimodal and bimodal oscillatory
regimes, as well as a SR that is observed close to the transition between the two regimes.

It is worth mentioning that, while GMM belongs to the class of
``non-linear'' models whose mean-field dynamics is associated to a
double-well Ginzburg-Landau potential, the VM with contrarians
described above belongs to a completely
different class characterized by an associated zero potential that
leads to a dynamics driven purely by noise \cite{Vazquez-2008-c}.  A main consequence of this difference is that the average magnetization
is conserved in the VM, while it is not in the GMM.  Another
consequence is that, in the version of these models with contrarians,
the order-disorder transition in the thermodynamic limit ($N \to \infty$) takes place
at a finite fraction of contrarian agents $a_c>0$ in the GMM, while
in the VM the transition happens at a vanishing contrarian probability ($p_c \to
0$).  We also need to mention that the SR effect has also
been observed in other opinion models.  For instance in
\cite{Gimenez-2012,Gimenez-2013} the authors found SR in a variation of
the Sznajd model with stochastic driving and a periodic signal.  The
work in \cite{Kuperman-2002} analyzed a majority rule dynamics under
the action of noise and an external modulation, and found a SR that
depends on the randomness of the small-world network.  There are also
other works \cite{Tessone-2005,Tessone-2009,Martins-2009,Muslim-2023,Mobilia-2023} that explored the combined effects of a stochastic driving and an external
signal on a majority rule dynamics.  However, none of these works have
incorporated a contrarian behavior in the dynamics.

The rest of the article is organized as follows.  We introduce the
model in section~\ref{model}.  In section~\ref{results} we present
numerical simulation results for the evolution of the system and the
behavior of different magnitudes that characterize the SR phenomena.
In section~\ref{mean-field} we develop a mean-field (MF) approach that
gives an insight into the system's evolution and the relation between the SR and the transition between different regimes.  Finally, in section~\ref{discussion} we summarize our findings and discuss the results.

\section{The~Model}
\label{model}

We consider a population of $N$ interacting agents where a given agent
$i$ ($i=1,..,N$) can hold one of two possible opinion states
$s_i=+1,-1$.  We denote by $\sigma_+(t)$ and $\sigma_-(t)$ the
fraction of nodes with respective states $+1$ and $-1$ at time $t$,
such that $\sigma_+(t)+\sigma_-(t)=1$ for all 
$t\ge 0$.  In a time step $\Delta t=1/N$ of the dynamics, we follow the basic GMM using groups of size three to update individual opinions. However, here for our purpose of investigating the effects of propaganda on individuals, we implement the rule in a different setting, which does not modify the outcome. Instead of selecting three agents randomly to update all of them at once, we pick one agent $i$ with state $s_i$ and a group of three other different agents $j, k, l$ ($i \ne j \ne k \ne l$), all randomly chosen.  In the $N \gg 1$ limit, their respective states are $(s_j, s_k, s_l)$ with probability $\sigma_{s_j} \sigma_{s_k} \sigma_{s_l}$.  A majority of $+$ choices is thus obtained for the configurations $(+,+,+)$, $(+,+,-)$, $(+,-,+)$ and $(-,+,+)$, yielding an overall probability
\begin{equation}
  P_+ \equiv \sigma_+^3+3 \sigma_+^2 \sigma_-.
  \label{P+}
\end{equation}
Similarly, a majority of $-$ occurs for $(-,-,-)$, $(+,-,-)$, $(-,+,-)$ and $(-,-,+)$, with the overall probability
\begin{equation}
  P_- \equiv \sigma_-^3 + 3 \sigma_-^2 \sigma_+.
  \label{P-}
\end{equation}
Then, agent $i$ updates its state in two steps.  i) First, the update
follows the basic GMM, where agent $i$ simply adopts the majority
state of the group of the three agents $j, k, l$.  We thus have $s_i
\to s_i=+1$ with probability $P_+$, or $s_i \to s_i=-1$ with
probability $P_-=1-P_+$.  ii) Second, agent $i$ can either preserve
this majority state ($s_i \to s_i$) with probability $p_{s_i}$, or
change to the opposite (minority) state ($s_i \to -s_i$) with the
complementary probability $1-p_{s_i}$, where $p_{s_i}$ is defined
below. The implication of this second step is that each agent can
behave as a "contrarian'' by adopting the state opposed to the majority
(minority state) with probability $1-p_{s_i}$, or as a "majority follower" with
probability $p_{s_i}$.  Thus, there is no fixed fraction of contrarian
agents as in \cite{Galam-2004}  

At this point, we introduce the effect of an external field $H$ on agent $i$ in state $s_i$ within a Boltzmann scheme, by assuming that the probability $p_{s_i}$ to preserve the majority state is larger when $s_i$ is aligned with $H$ [i.e. $\mbox{sign}(s_i) = \mbox{sign}(H)$], 

\begin{equation}
p_{s_i,H}=\frac{e^{\left[s_i H\right]/T}}{e^{\left[s_i H\right]/T} + e^{-\left[s_i H\right]/T}} \;,
\label{pH} 
\end{equation}
where $T \ge 0$ is a parameter that plays the role of a \emph{social temperature} analogous to the contrarian feature of the GMM.  The related probability to oppose the field is $1-p_{s_i,H}$.  We assume that $H$ is an oscillating periodic field $H(t)=H_0 \sin(\omega t)$ with amplitude $H_0$ ($0 \le H_0 \le 1$), frequency $\omega=2 \pi/\tau$ and period $\tau$, which represents an external propaganda.  Thus, according to Eq.~(\ref{pH}), agents are more likely to keep the opinion that is aligned with the propaganda.  In addition to the external field, we introduce an individual ``inertia'' parameter $I$, which provides an agent with a weight to preserve its current state against a field favoring the opposite state.  It is a self-interaction $-I s_i s_i$ that modifies Eq.~(\ref{pH}) as
\begin{equation}
p_{s_i,a,H} = \frac{e^{\left[I s_i +H\right] s_i /T}}{e^{\left[I s_i +H\right]s_i/T} + e^{-\left[I s_i +H\right] s_i/T}} \;,
\label{paH} 
\end{equation}  
which can be rewritten as 
\begin{equation}
p_{s_i,1,H} = \frac{e^{\left[1+s_i H\right]/T}}{e^{\left[1+s_i H\right]/T} + e^{-\left[1+s_i  H\right]/T}} \;,
\label{p1H} 
\end{equation}  
where $I, H, T$ have been rescaled as $1, \frac{H}{I}, \frac{T}{I}$ using $s_i^2=1$.

At this stage we combine the GMM with the inertia and field effects by taking
\begin{equation}
p_{s_i}(t)= \frac{e^{\left[1+s_i  H(t) \right]/T}}{e^{\left[1+s_i  H(t) \right]/T} + e^{-\left[1+s_i  H(t) \right]/T}}
\label{psi} 
\end{equation}  
for the probability of agent $i$ to keep the majority state $s_i$, and $1-p_{s_i}(t)$ for the probability to adopt the opposite (minority) state $-s_i$, which can be interpreted as a noise. Finally, combining Eqs.~(\ref{P+}), (\ref{P-}) and (\ref{psi}), the probability $\mathcal P_+$ for a randomly selected agent $i$ to adopt the state $+$ in a single time step $\Delta t$ is given by
\begin{eqnarray}
\mathcal P_+ = (\sigma_+^3+3 \sigma_+^2 \sigma_-) \frac{e^{\left[1+H(t) \right]/T}}{e^{\left[1+ H(t) \right]/T} + e^{-\left[1+  H(t) \right]/T}} 
+ (\sigma_-^3+3 \sigma_-^2 \sigma_+)\frac{e^{-\left[1-H(t) \right]/T}}{e^{\left[1-H(t) \right]/T} + e^{-\left[1-  H(t) \right]/T}} \;,
\label{p+}
\end{eqnarray}
where the first term comes from following a local majority $+$ among the three selected agents, which happens with probability $P_+ p_+(t)$, while the second term corresponds to opposing the state $-$ in case of a majority of $-$ among the three selected agents, which happens with probability $P_- [1-p_-(t)]$.  Analogously, the state $-$ is selected with probability $\mathcal P_- \equiv 1-\mathcal P_+$.

As noted above, only the ``focal agent'' $i$ updates its state, unlike in the original GMM where all agents in the chosen group update their states. Equation~(\ref{psi}) shows that individuals are more prone to adopt the opinion of the majority when it is aligned with the propaganda.  In addition, $p_+$ and $p_-$ approach the value $1$ as $T \to 0$, which makes this case equivalent to the original GMM, with neither contrarians nor external field. In the opposite limit $T \to \infty$, $p_+$ and $p_-$ approach the value $1/2$, which corresponds to the pure noise case where agents take one of the two opinions at random, independent of the field.

\section{Numerical results}
\label{results}

\subsection{Evolution of the magnetization}
\label{magnetization}

\begin{figure}[t]
  \begin{center}
    \includegraphics[width=6.0cm]{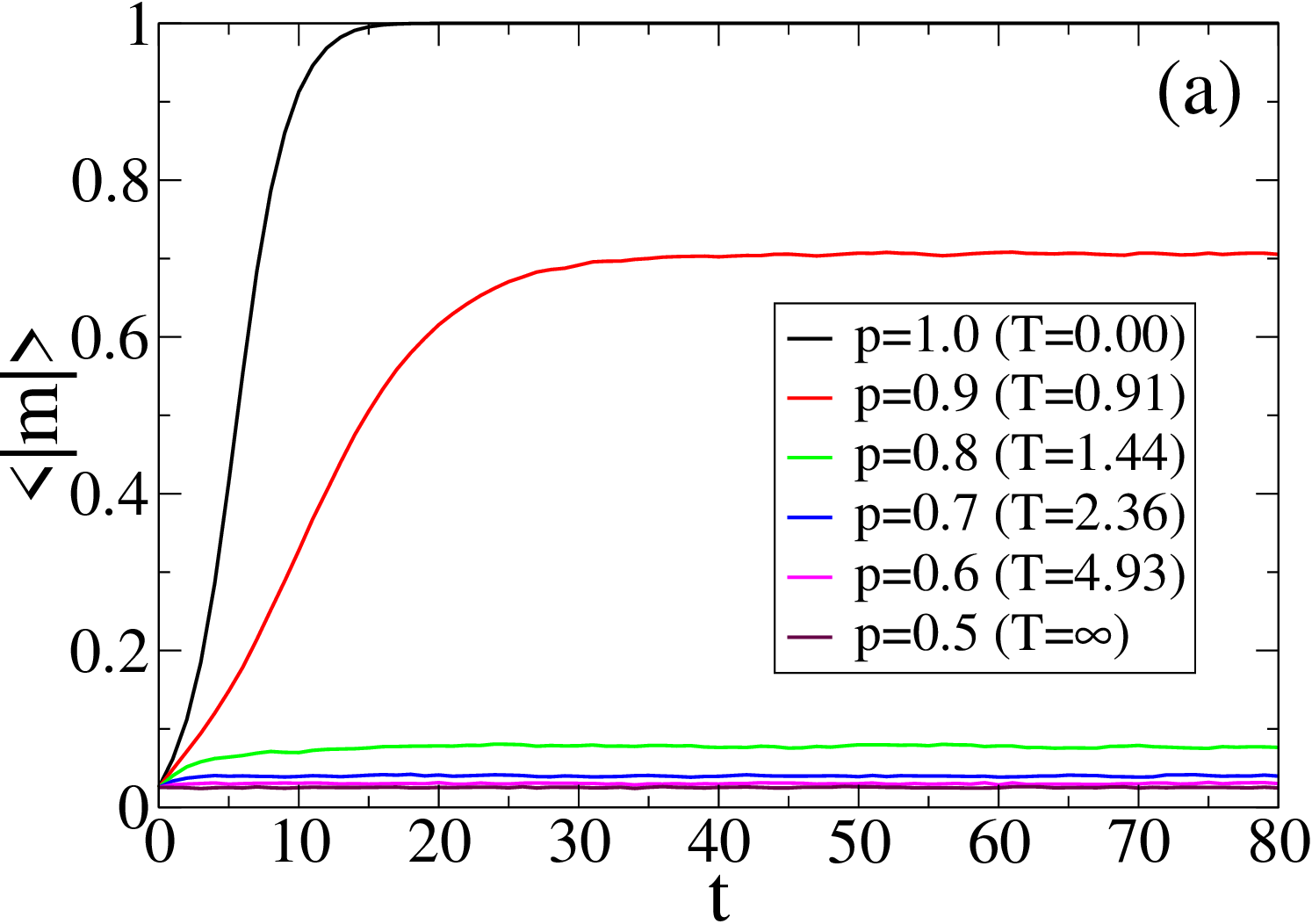} ~~~~~
    \includegraphics[width=6.0cm]{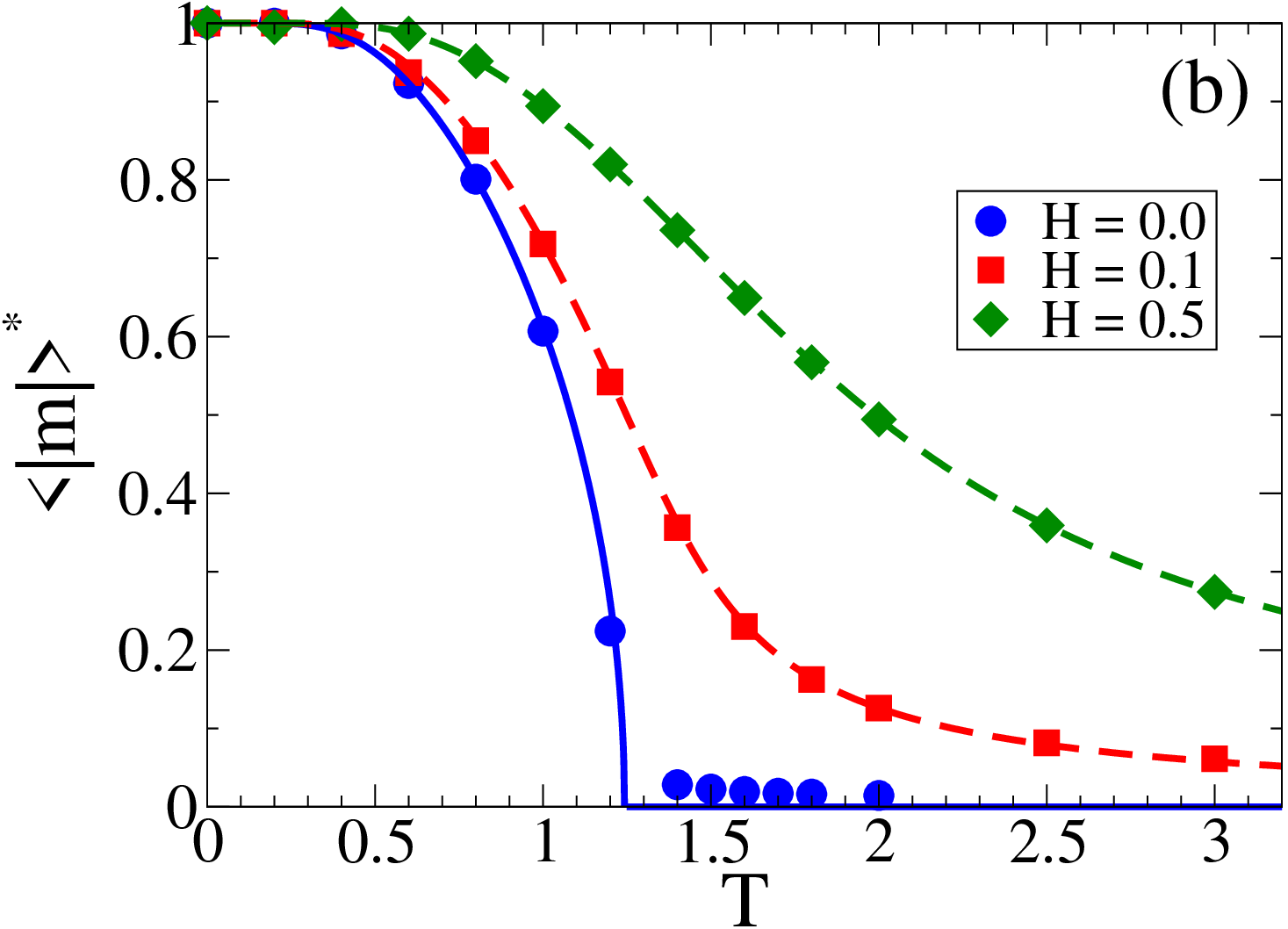}
  \end{center}  
  \caption{(a) Time evolution of the average value of the absolute
    magnetization $|m|$ in a population of $N=10^3$ agents, zero field
    $H=0$ and various values of majority probability
    $p=(1+e^{-2/T})^{-1}$, as indicated in the legend.  (b) Stationary
    value of $\langle |m| \rangle$ vs $T$ for constant fields $H=0.0$
    (circles), $H=0.1$ (squares) and $H=0.5$ (diamonds).  The solid line is the analytical expression from Eq.~(\ref{m-t-H-0}), while the dashed lines are the numerical integration of Eq.~(\ref{dmdt}).  The averages were done over $10^3$ independent realizations starting from a symmetric condition $m_0=0$.}
  \label{m_ave}
\end{figure}

We start by studying the time evolution of the mean opinion of the population or magnetization defined as $m(t) \equiv \frac{1}{N}\sum_{i=1}^N s_i(t)$, 
for the simplest case of zero field $H=0$, which corresponds to the contrarian GMM with symmetric majority probabilities $p_+=p_-=p=(1+e^{-2/T})^{-1}$.  
We run several independent realizations of the dynamics where, initially, 
each agent adopts state $+1$ or $-1$ with respective probabilities 
$\sigma_+(0)$ and $\sigma_-(0)$, leading to an initial average magnetization 
$m(0)=\sigma_+(0)-\sigma_-(0)$.  Due to the symmetry of the system, 
the evolution of the average value of $m$ over many realizations starting 
from $m(0)=0$ gives $\langle m \rangle(t) \simeq 0$ for all $t \ge 0$, which does not describe the correct behavior of the system.  
Instead, we looked at the evolution of the absolute value of the magnetization, $|m|$, 
as we show in Fig.~\ref{m_ave}(a), for various values of $p$.  In Fig.~\ref{m_ave}(b) we show in circles the stationary value of $\langle |m| \rangle$ as a function of $T$ for $H=0$.  We observe that, as $T$ increases, the system displays a transition from an ordered state ($|m|>0$) for $T<T_c^0$, to a disordered state ($|m| \simeq 0$) for $T>T_c^0$, where $T_c^0$ is a transition temperature.  This order-disorder transition, reminiscent of the GMM with a fixed fraction of contrarian agents \cite{Galam-2004}, is induced by the presence of a contrarian behavior that acts as a source of external noise, preventing the system to reach full consensus.  When the noise amplitude, controlled by $T$, overcomes a threshold value $T_c^0$ the system reaches complete disorder.  In section~\ref{mean-field} we develop a mean-field approach that allows to estimate the transition temperature as $T_c^0 \simeq 1.24$.  When the field is turned on, these results change completely.  In the case that the field remains constant in time (constant propaganda H=\mbox{const}),
the symmetry of the system is broken in direction of $H$, increasing
the stationary value of $\langle |m| \rangle$ as compared to the $H=0$
case.  This effect can be seen in Fig.~\ref{m_ave}(b), where we see
that $\langle |m| \rangle^*$ increases monotonically with $H$.
Besides, the order-disorder transition disappears for $H>0$ (see
$H=0.1$ and $H=0.5$ curves).  \\

\begin{figure}[t]
  \begin{center}
    \includegraphics[width=4.5cm]{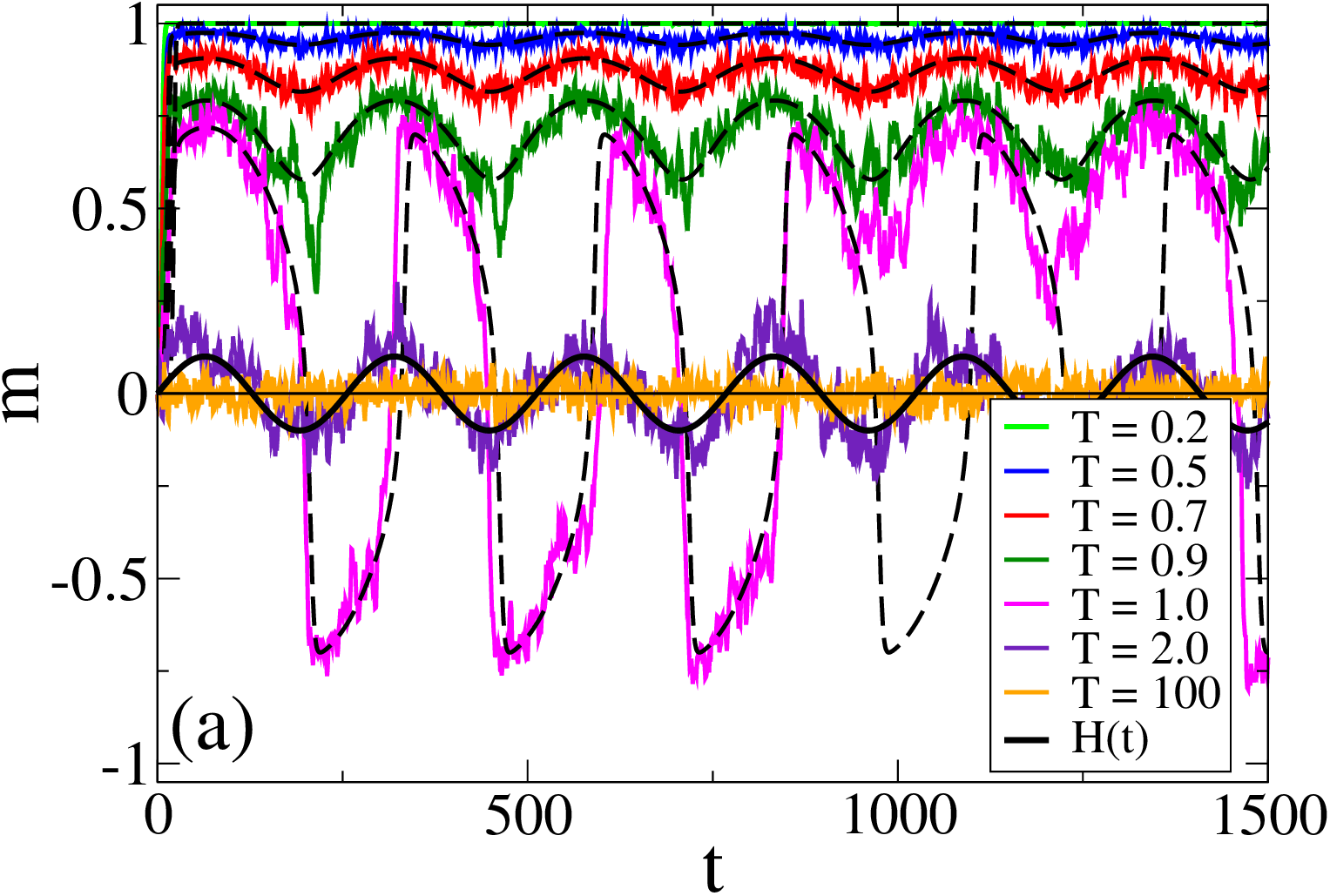}
    \includegraphics[width=4.5cm]{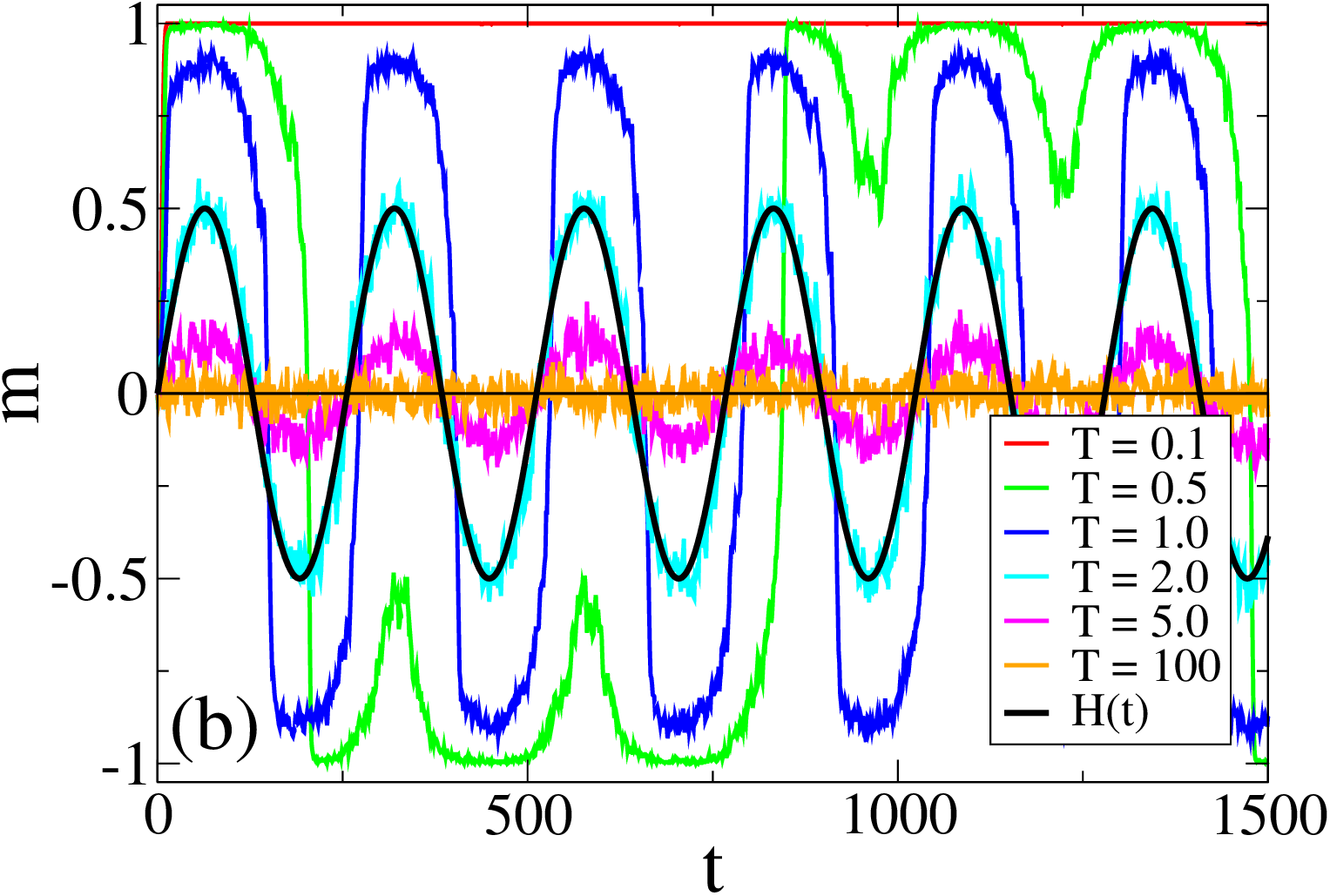}    
    \includegraphics[width=4.5cm]{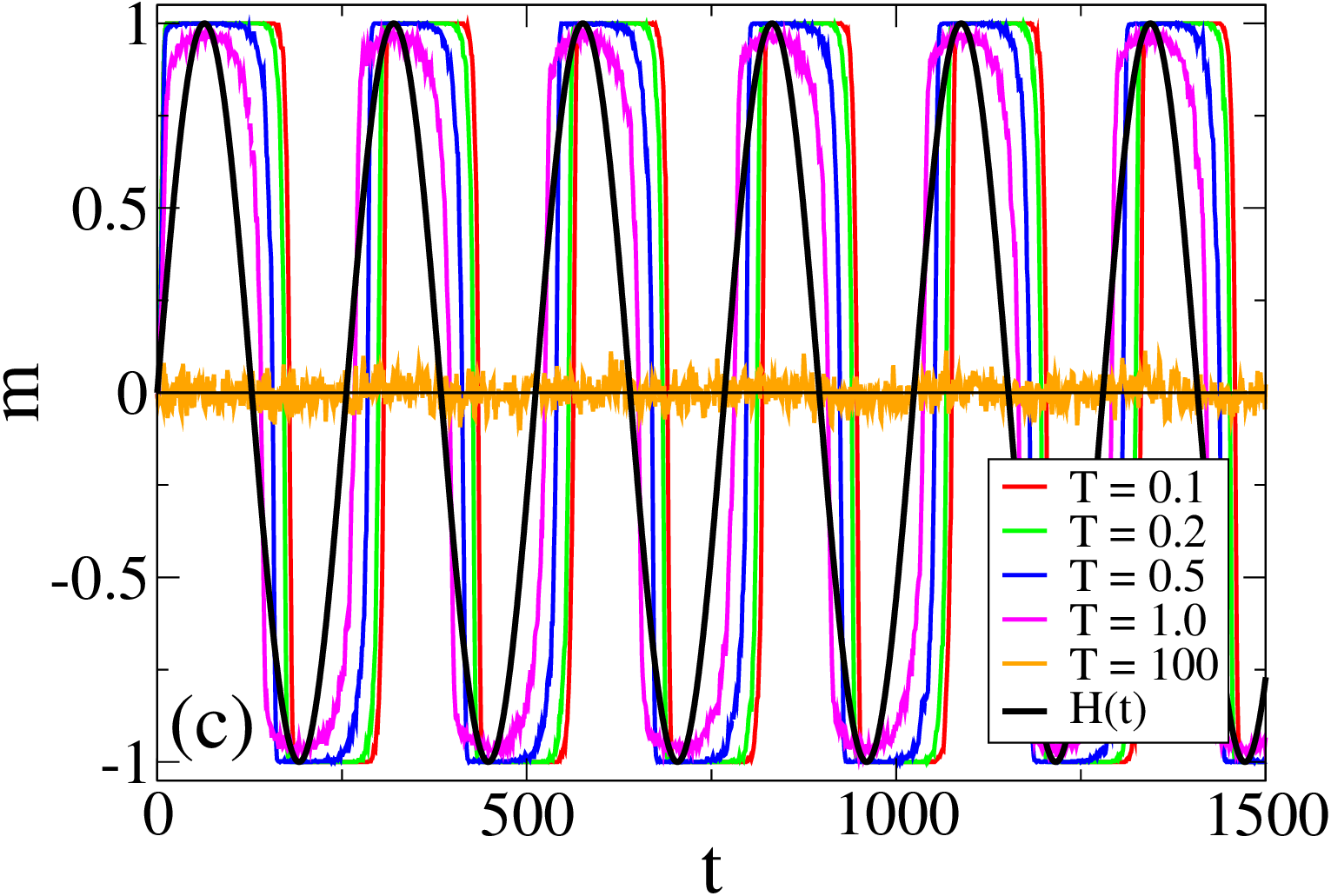}     
  \end{center}  
  \caption{Evolution of $m$ in a single realization for a population of $N=1024$ 
  agents under an oscillating field with period $\tau=256$ and amplitudes 
  $H_0=0.1$, $0.5$ and $1.0$, panels (a), (b) and (c), respectively, 
  and the temperatures indicated in the legends.  Solid lines correspond to MC simulations, while dashed lines in panel (a) represent the numerical integration of Eq.~(\ref{dmdt}).}
  \label{m-t-256}
\end{figure}

If we now let the field oscillate in
time, a series of different regimes emerge.  In Fig.~\ref{m-t-256} we
show the evolution of $m$ in a single realization under the effects of
an oscillating field, for three different amplitudes $H_0$, period
$\tau=256$ and various temperatures.  For the $H_0=0.1$ and $H_0=0.5$
cases [panels (a) and (b)], we can see that for low temperatures $m$
oscillates around a positive value or negative value, and that oscillations vanish for small enough $T$, where $m$ stays in a value close to $1.0$ (consensus), as we can see for $T=0.2$ and $T=0.1$ in panels (a) and (b), respectively.  The center of oscillations can jump from positive to negative values and vice-versa (bimodal regime), as we can see in panel (b) for $T=0.5$.  Above a given temperature threshold, $T_c \simeq 1.0$ for $H_0=0.1$ [panel (a)] and $T_c \simeq 0.5$ for $H_0=0.5$ [panel (b)], the magnetization oscillates around $m=0$ (unimodal regime).  This behavior is reminiscent of the ordered and disordered phases in the model without field [Fig.~\ref{m_ave}(b)], although the transition temperature $T_c^0 \simeq 1.24$ for $H=0$ is quite different from that of the model with oscillating field.  An insight into this behavior shall be given in section \ref{mean-field}.  For $H_0=1.0$ [panel (c)] oscillations are centered at $m=0$ even for small $T$, and thus the bimodal regime is not observed.  Finally, at very large temperatures the high levels of noise leads to a purely stochastic dynamics where agents adopt an opinion at random, and thus $m$ fluctuates around zero.

\subsection{Residence times}
\label{residence}

\begin{figure}[t]
  \begin{center}
    \includegraphics[width=6.5cm]{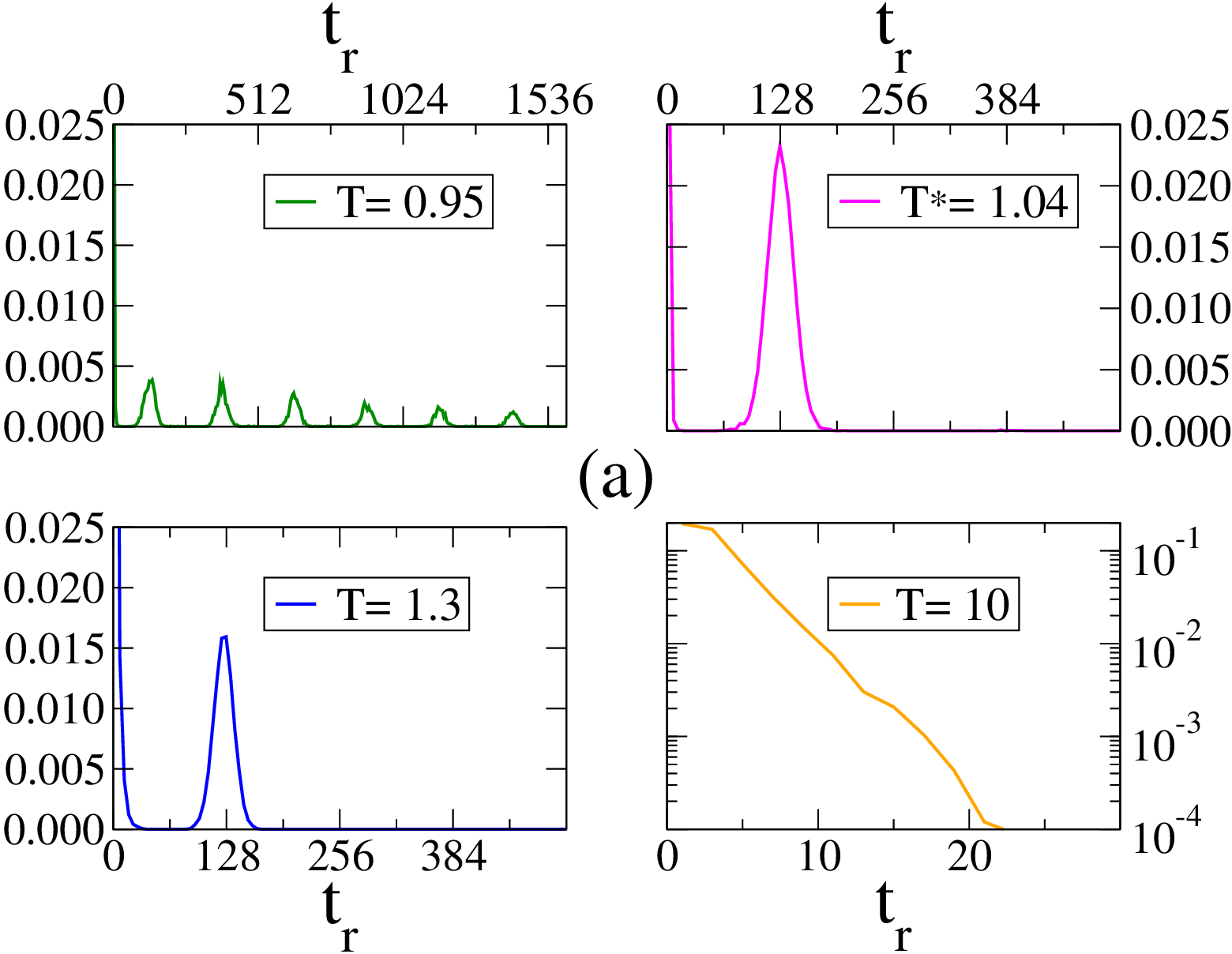} ~~~~~~
    \includegraphics[width=6.5cm]{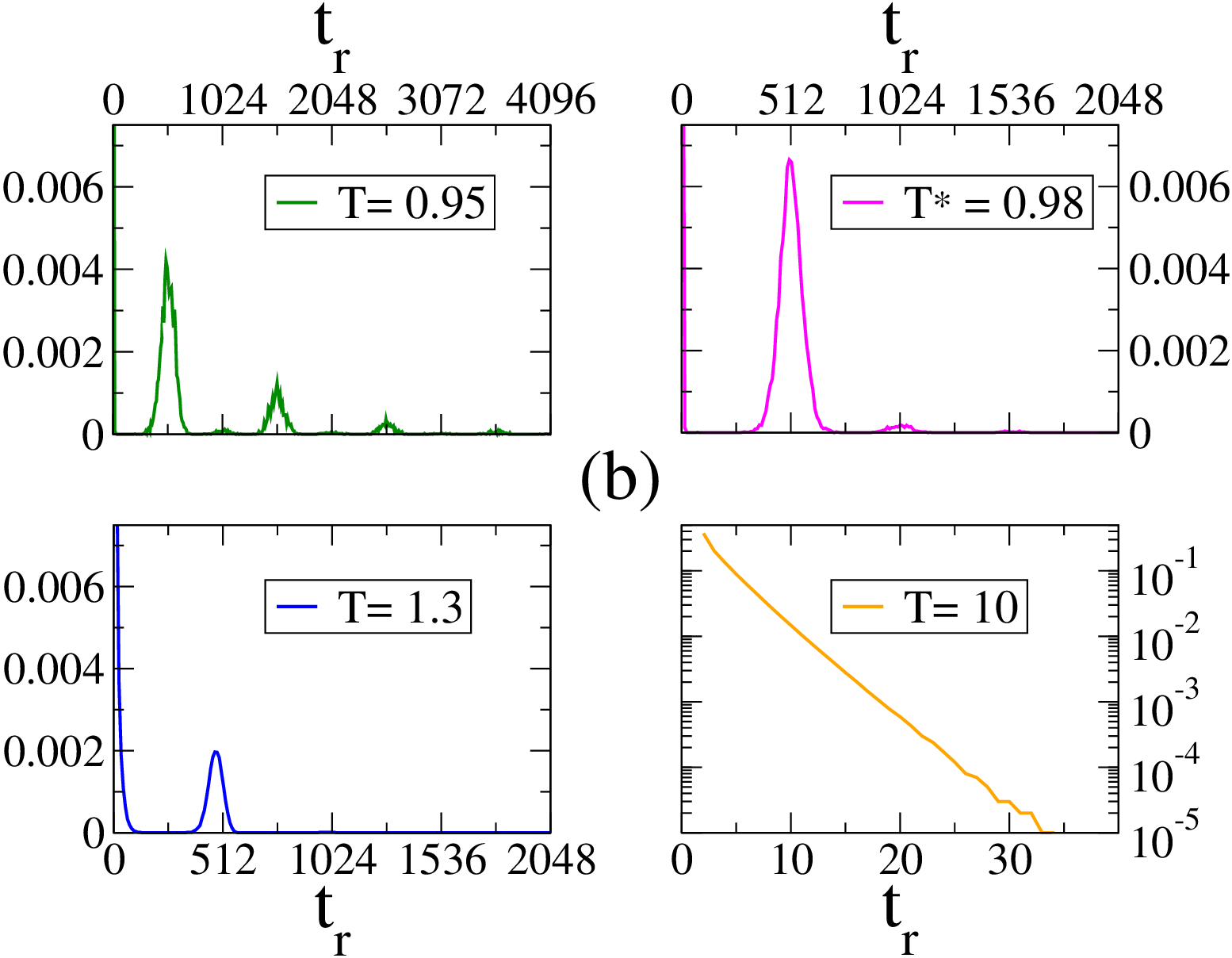}
  \end{center}  
  \caption{Normalized histograms of the residence time $t_r$ in a system of
    $N=1025$ agents under a field of amplitude $H_0=0.1$, period
    $\tau=256$ (a) and $\tau=1024$ (b), and the temperatures indicated
    in the legends.  The bottom-right panels are on a linear-log scale.} 
  \label{rtd}
\end{figure}

In order to characterize the different regimes described in the last
section, we study here the residence time $t_r$, defined as the time
interval between two consecutive changes of the sign of $m$, i.e.,
when $m$ crosses the center value $m=0$. In a single realization, $m$
can change sign multiple times depending on the parameter values,
leading to a distribution of the residence time that is particular of
each regime.  Results are shown in Fig.~\ref{rtd} for $N=1025$,
$H_0=0.1$, $\tau=256$ [panel (a)] and $\tau=1024$ [panel (b)].  In the
unimodal regime $m$ follows the oscillations of $H(t)$ around zero, and thus
$m$ tends to change sign when $H$ does, every time interval $\tau/2$.
Therefore, the residence time distribution ($RTD$) is peaked at $t_r \simeq
\tau/2$, as shown in panel (a) for temperatures $T=1.04$ and $T=1.3$, and in panel (b) for $T=0.98$ and $T=1.3$.  In the bimodal regime, the
RTD exhibits multiple peaks at $t_r=(n+1/2)\tau$ ($n=0,1,2,..$) (see
panels for $T=0.95$).  Here $m$ tends to perform oscillations around a
positive (negative) value until it changes to negative (positive)
oscillations, and back to positive (negative) oscillations again, as
we observe in Fig.~\ref{m-t-256}(b) for $T=0.5$. These changes are
more likely to happen when $H$ changes sign, in the first attempt at
time $t=\tau/2$, or in the second attempt one period later (at
$t=3\tau/2$), or in the third attempt at $t=5\tau/2$ and so on,
leading to the different peaks in the $RTD$.  Finally, for very large $T$ the $RTD$ shows an exponential decay due to the stochastic fluctuations of $m$ around zero (panels for $T=10$).

\subsection{Stochastic resonance}
\label{resonance}
 
\begin{figure}[t]
  \begin{center}
    \includegraphics[width=6.5cm]{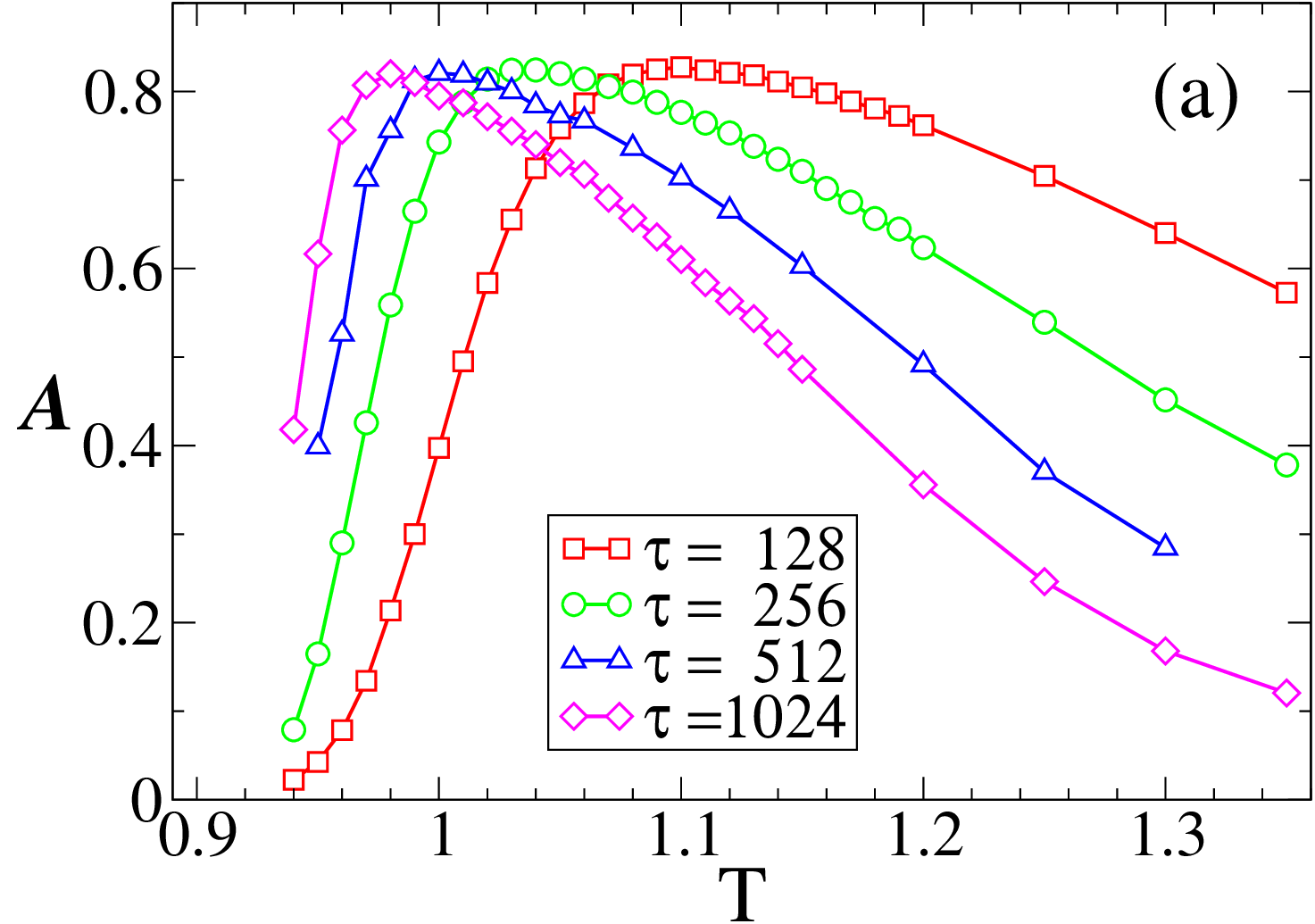} ~~~~~
    \includegraphics[width=6.5cm]{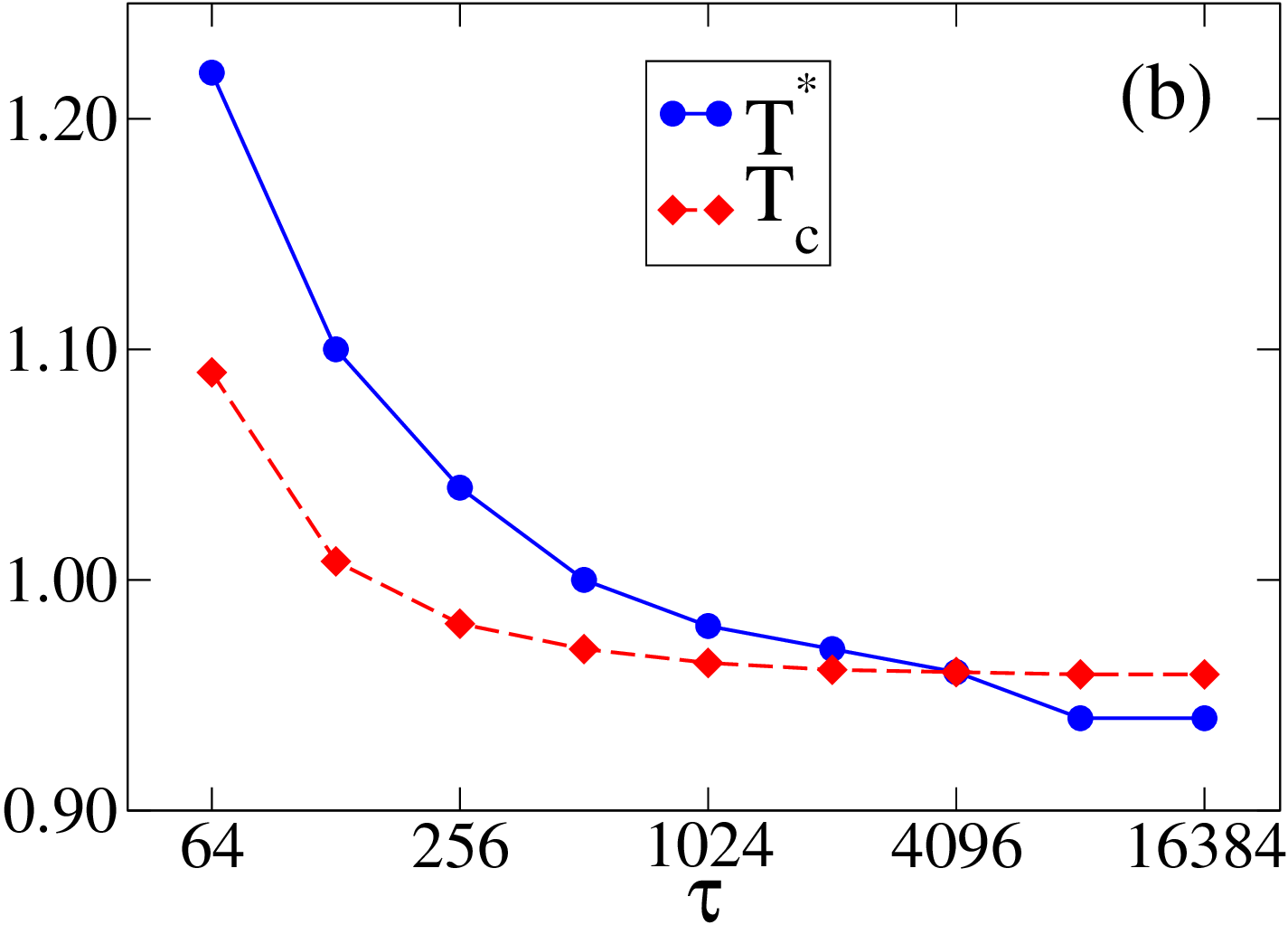}    
  \end{center}  
  \caption{(a) Response $\mathcal A$ as a function of the temperature $T$ for
    a field of amplitude $H_0=0.1$ and periods $\tau$ indicated in the legend.  (b) Resonance temperature $T^*$ [maximum of $\mathcal A$ vs $T$ curves from (a)] and transition temperature $T_c$ vs period $\tau$.}
  \label{Area} 
\end{figure}

The patterns of the RTD shown in section~\ref{residence} can be
employed to quantify the phenomena of stochastic resonance, as it was
done in related systems \cite{Gammaitoni-1998,Kuperman-2002}.  The
sensitivity or response of the system to the external oscillating
field can be measured by the area $\mathcal A$ under the first peak around
$\tau/2$ in the RTD histogram. It is expected that $\mathcal A$
reaches a maximum at the resonance temperature $T^*$, when $m$
resonates with the field $H$.  This method to quantify the resonance
is an alternative to the study of the signal-to-noise ratio
\cite{Gimenez-2012,Gimenez-2013,Gimenez-2022}.  Figure~\ref{Area}(a)
shows the response $\mathcal A$ vs $T$ for a field of amplitude
$H_0=0.1$.  Each curve corresponds to a different period $\tau$.  We
observe that $\mathcal A$ reaches a maximum value at a temperature $T^*$
that depends on $\tau$.  The RTD for the resonance temperatures
$T^*=1.04$ and $T^*=0.98$ for periods $\tau=256$ and $\tau=1024$,
respectively, are shown in the top-right panels of Figs.~\ref{rtd}(a)
and \ref{rtd}(b), where we see the existence of a well defined peak
centered at $t_r=\tau/2$.  For larger temperatures (see $T=1.3$) 
there is also a peak at $\tau/2$, although lower than that for $T^*$, and the RTD
exhibits another pronounced peak near $t_r=0$, corresponding to the short crossings of
$m(t)$ that become more frequent as $T$ increases (larger fluctuations in
$m$). 
     
\section{Mean-field approach}
\label{mean-field}

In this section we analyze the behavior of the model within a MF approach, by deriving a rate equation for the evolution of $m$ that corresponds to the dynamics introduced in section~\ref{model}.  Let us write the fractions of $+$ and $-$ agents in terms of the magnetization $m$, $\sigma_+=(1+m)/2$ and $\sigma_-=(1-m)/2$.  As we described in section~\ref{model}, in a time step $\Delta t=1/N$ a random agent $i$ with state $s_i=-1$ is chosen with probability $\sigma_-$, and then adopts the state $+$ ($s_i=-1 \to s_i=+1$ flip) with probability $\mathcal P_+=P_+ p_+ + P_- (1-p_-)$, which corresponds to adopt either the majority state $+$ of a selected $+$ majority, or the minority state $+$ of a selected $-$ majority, where $P_+$ and $P_-$ are given by Eqs.~(\ref{P+}) and (\ref{P-}), respectively.  This flip $-1 \to +1$ leads to an overall change $\Delta m=2/N$ in $m$.  Conversely, with probability $\sigma_+$ the chosen agent $i$ has state $+1$, and flips to $-1$ ($s_i=+1 \to s_i=-1$ flip) with probability $\mathcal P_- = P_- p_- + P_+ (1-p_+)$, leading to a change $\Delta m=-2/N$.  Assembling these factors, the mean change of $m$ in a time step can be written as 
\begin{eqnarray*}
	&& \frac{dm}{dt} =  \frac{1}{1/N} \left[ \sigma_- \mathcal P_+ \left( \frac{2}{N} \right) - \sigma_+ \mathcal P_-  \left( \frac{2}{N} \right) \right],
\end{eqnarray*}
which becomes, in the $N \to \infty$ limit, the rate equation
\begin{eqnarray}
	\frac{dm}{dt} = \frac{1}{2} m (m^2-5) + \frac{1}{2} p_+ (1+m)^2 (2-m) - \frac{1}{2} p_- (1-m)^2 (2+m),
	\label{dmdt}
\end{eqnarray}
after replacing the expressions for $\mathcal P_+$ and $\mathcal P_-$ and doing some algebra.  Here 
\begin{equation}
	p_+(t) = \frac{e^{\left[1+ H(t)\right]/T}}{e^{\left[1+ H(t)\right]/T} + e^{-\left[1+ H(t)\right]/T}} ~~~ \mbox{and} ~~~  
	p_-(t) = \frac{e^{\left[1- H(t)\right]/T}}{e^{\left[1- H(t)\right]/T} + e^{-\left[1- H(t)\right]/T}}
\end{equation}  
are the probabilities of adopting the state $+1$ and $-1$ of a majority, respectively, as defined in Eq.~(\ref{psi}). \\

For the zero field case ($H_0=0$) is $p_+=p_-=p=(1+e^{-2/T})^{-1}$, and thus Eq.~(\ref{dmdt}) is reduced to the simple equation
\begin{eqnarray}
  \frac{dm}{dt} = \frac{1}{2} m \left[ 6p-5-(2p-1)m^2 \right].
  \label{dmdt-1}
\end{eqnarray}
Equation~(\ref{dmdt-1}) has three fixed points corresponding to the possible stationary states of the agent based model.  The fixed point $m_0^*=0$ is stable for $p<5/6$ and corresponds to a disordered active state with equal fractions of $+$ and $-$ agents ($\sigma_+=\sigma_-=1/2$), whereas the two fixed points
\begin{equation}
  m_{\pm}^*=\pm \sqrt{\frac{6p-5}{2p-1}}
  \label{m-t-H-0}
\end{equation}
are stable for $p>5/6$, and they represent asymmetric active states
of coexistence of $+$ and $-$ agents, with stationary fractions
$\sigma_+^*=(1 + m_+^*)/2 > \sigma_-^*=(1 - m_+^*)/2$ and
$\sigma_+^*=(1 + m_-^*)/2 < \sigma_-^*=(1 - m_-^*)/2$.  The stable fixed points are plotted by a solid line in Fig.~\ref{m_ave}(b), where we observe a good agreement with MC simulation results (solid circles).  Equation~(\ref{m-t-H-0}) shows the existence of a transition from order to disorder as $T$ overcomes the value $T_c^0=2/\ln(5) \simeq 1.24$ ($p_c^0=5/6$), as we already mentioned in section~\ref{magnetization}.  Notice that the probability of behaving as a contrarian $1-p_c^0=1/6$ is identical to the critical proportion of contrarians $a_c=1/6$ obtained in the GMM for groups of size $3$ \cite{Galam-2004}.  Given that Eq.~(\ref{dmdt-1}) can be rewritten as a Ginzburg-Landau equation with an associated double-well potential with two minima at $m_{\pm}^*$, we expect a bistable regime for $T<T_c$, where in a single realization $m$ jumps between $m_+^*$ and $m_-^*$. \\

For a field that is constant in time ($H=\mbox{const} \ne 0$) the fixed points of Eq.~(\ref{dmdt}) are given by the roots of a cubic polynomial, and $m=0$ is not longer a root.  Only one root is real, and corresponds to the stationary state of the agent-based model.  As the analytical expression for the real root is large and not very useful, we integrated Eq.~(\ref{dmdt}) numerically to find the stationary value $m^*$, which we plot by a dashed line in Fig.~\ref{m_ave}(b) for $H=0.1$ and $0.5$.  We observe a good agreement with MC simulations (symbols).  A positive field $H>0$ breaks the symmetry in favor of the $+$ state, given that $p_+>p_-$, leading to a positive stationary value $m^*>0$ that increases monotonically with $H$. \\

\begin{figure}[t]
  \begin{center}
    \includegraphics[width=6.0cm]{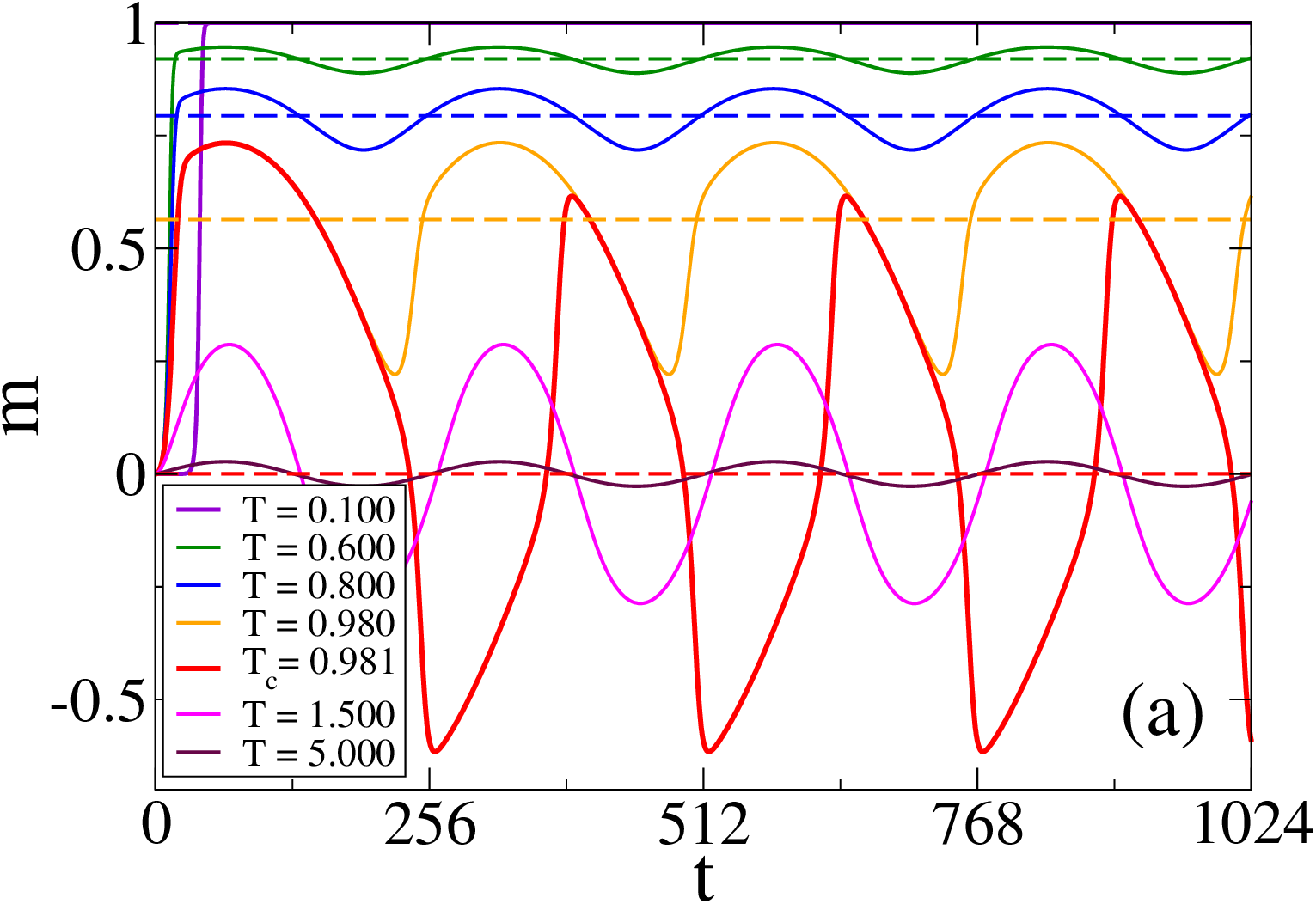} ~~~~	
    \includegraphics[width=6.0cm]{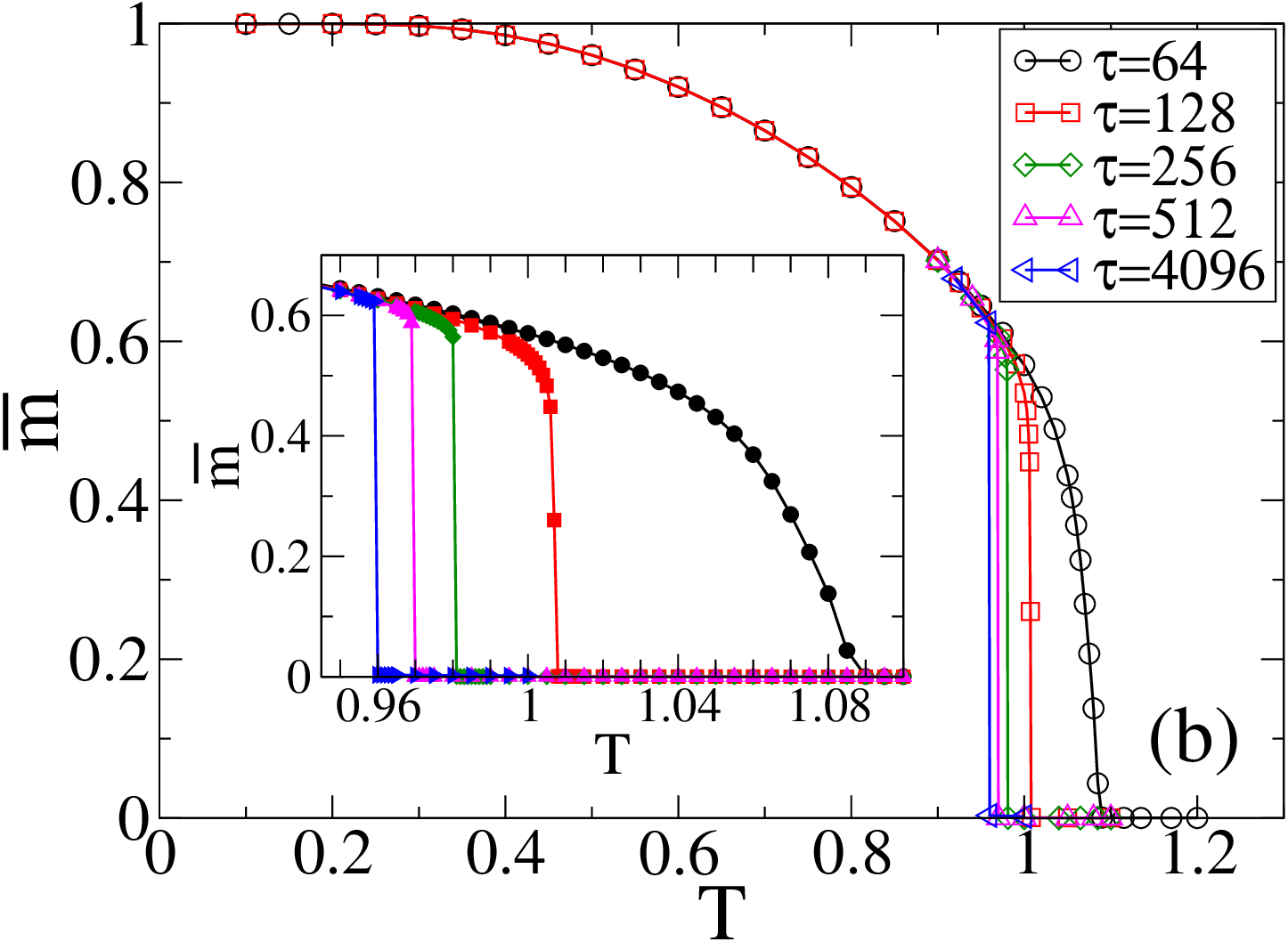} 
  \end{center}  
  \caption{(a) Time evolution of the magnetization $m$ from Eq.~(\ref{dmdt}) for a field of amplitude $H_0=0.1$, period $\tau=256$, and the temperatures indicated in the legend.  Horizontal dashed lines represent the time average value of $m$, $\overline{m}$, in the interval $t \in (0,1000 \tau)$.  (b) Time average of the magnetization, $\overline{m}$, vs temperature $T$ for the field's periods indicated in the legend.  The inset shows a closer look around the transition values $T_c$.}
  \label{m-ave-T}
\end{figure}

For an oscillating field $H(t)$, we have that $p_+(t)$ and $p_-(t)$ oscillate in time according to $H(t)$, which in turn leads to oscillations in $m(t)$.  In order to explore, within the MF approach,  the behavior of $m$ in the different regimes described in section~\ref{magnetization}, we plot in Fig.~\ref{m-ave-T}(a) the evolution of $m$ obtained from the numerical integration of Eq.~(\ref{dmdt}) for $H_0=0.1$, $\tau=256$, and various temperatures.  For low temperatures we see that $m$ oscillates around a positive value (it could also be a negative value for other initial conditions), but when the temperature is increased beyond a threshold value oscillations turn to be around $m=0$.  At first sight, this transition that happens in the oscillatory regime of $m$, already reported in section~\ref{magnetization} from MC simulations, appears to be quite sharp, where the center of oscillations seems to jump from a large value to zero after a small increment of $T$.  To better characterize the transition we plot in Fig.~\ref{m-ave-T}(b) the temporal average of $m$ from $t=0$ to $t=1000 \tau$, called $\overline{m}$, as a function of $T$ and for several periods $\tau$.  The value of $\overline{m}$ can be seen as an order parameter, which takes a positive or negative value in the bimodal regime and a value close to zero in the unimodal regime.  We can see that $\overline{m}$ decreases continuously with $T$ for low $\tau$ (see curve for $\tau=64$), and that the transition becomes more abrupt as $\tau$ increases (see curves for $\tau \ge 256$).  The inset shows a more detailed view of the transition in the value of $\overline{m}$. 

In Fig.~\ref{m-t-256}(a) we compare the evolution of $m$ obtained from the MF approach (dashed lines) with that from MC simulations, for $H_0=0.1$, $\tau=256$, and various temperatures.   We observe a good agreement with single realizations of the dynamics, except for the temperature $T=1.0$ that is close to the transition value $T_c \simeq 0.981$, estimated from Fig.~\ref{m-ave-T}(b) as the point where $\overline{m}$ becomes zero.  This discrepancy is due to the fact that the MF approach cannot reproduce the random jumps of $\overline{m}$ from the value $\overline{m} \simeq 0.564$ in the bimodal regime to $\overline{m} \simeq 0$ in the unimodal regime.  These jumps are induced by finite-size fluctuations, and are more frequent when the control parameter $T$ is close to the transition point $T_c$. \\

An insight into the behavior of the resonance temperature $T^*$ with the period $\tau$ can be obtained from the MF approach assuming that the response $\mathcal A$ reaches a maximum value at a temperature similar to the transition point $T_c$, that is, we expect $T^* \simeq T_c$.  This is because in the bimodal regime $T<T_c$ the magnetization $m$ oscillates around a positive or a negative value and eventually crosses $m=0$ around times $t=\tau/2$, $3\tau/2$, etc., by finite-size fluctuations, leading to multiple peaks in the residence time distribution.  Then, at $T=T_c$, oscillations start to be centered at $\overline{m}=0$, and thus we expect that the $RTD$ shows a single peak at $\tau/2$.  By increasing $T$ beyond $T_c$ we expect that the height of the peak for $T=T_c$ is reduced by the presence of a higher noise that induces another maximum of the $RTD$ at $t=0$, as explained in section~\ref{residence}, leading to a smaller $\mathcal A$.  Therefore, we expect that $\mathcal A$ is maximum at $T \simeq T_c$.  Figure~\ref{Area}(b) shows in diamonds the value of $T_c$ obtained from Fig.~\ref{m-ave-T}(b) for various periods $\tau$.  We see that $T_c$ decreases with $\tau$, as it happens with $T^*$ (circles), although discrepancies between $T_c$ and $T^*$ increase as $\tau$ decreases.

\section{Summary and discussion}
\label{discussion}

In this article we studied the dynamics of the binary--state majority rule model introduced by Galam for opinion formation, under the presence of an external propaganda and contrarian behavior. When an agent has to update its opinion, it can either follow the majority opinion among three random neighbors, similarly to the original GMM, or adopt the opposite (contrary) opinion, i.e., the minority opinion.  The probability to adopt the majority opinion $p_\pm(t)$ is coupled to an external field that oscillates periodically in time (propaganda), in a way that agents are more likely to adopt the majority opinion when it is align with the field.  This rule tries to reproduce a reinforcing mechanism by which individuals have a tendency to follow the majority opinion when it is in line with mass media propaganda.  Besides, the majority probability $p_\pm$ depends on a parameter $T$ (temperature) that acts as an external source of noise, in such a way that by increasing $T$ from zero the system goes from following the majority opinion only ($p_\pm =1$ for $T=0$) to adopting a random opinion for large temperatures ($p_\pm=0.5$ for $T \gg 1$).

We explored the model in complete graph (all-to-all interactions) and found different phenomena associated to different regimes as $T$ is varied. For $T$ below a threshold value $T_c$ the system is in a bimodal regime, where the mean opinion $m$ oscillates in time around a positive or negative value, $\overline{m}_\pm$, and performs jumps between $\overline{m}_+$ and $\overline{m}_-$ due to finite-size fluctuations, similarly to what happens in a bistable system.  As the temperature is increased beyond $T_c$ there is a transition to an unimodal regime in which $m$ oscillates around zero, where the amplitude of oscillations decreases with $T$ and eventually vanishes in the $T \gg 1$ limit that corresponds to pure noise.  The transition at $T_c$ becomes more abrupt as the period $\tau$ of the field increases. We also studied the response of the system to the external field, by means of the distribution of residence times, i.e., the time interval between two consecutive changes of the sign of $m$.  We found that there is an optimal temperature $T^*$ for which the response is maximum, that is, a stochastic resonance phenomenon induced by the external noise controlled by $T$.  Also, we developed a mean-field approach that lead to a non-linear rate equation for the time evolution of $m$ in the thermodynamic limit, whose numerical solution agrees very well with MC simulations of the model.  We used this equation to give a numerical estimate of $T_c$, and found that the behavior of $T_c$ with the period $\tau$ is qualitatively similar to that of $T^*$.  Although the transition temperature $T_c$ is similar to the resonance temperature $T^*$ only for large $\tau$, this analysis shows that they are related.

A possible interpretation of these results in a social context is the following.  Reacting with a contrarian attitude occasionally (small $T$/low noise) on a given issue, that is, adopting an opposite position to that of the majority of our acquaintances, leads to a state of collective agreement in a population, which can be reversed completely after some time by means of a collective decision, independently of the external propaganda.  This alternating behavior between opposite opinions might be seen as more "socially healthy" than a frozen full consensus in one of the two alternatives, which happens in populations with a total absence of contrarian attitudes ($T=0$). However, having a contrarian behavior more often could induce a collective state where the mean opinion oscillates in time following the external propaganda, which can be interpreted as a society whose opinions are manipulated optimally by the mass media, in opposition to collective freedom.  Finally, in the extreme case of having a very frequent contrarian attitude ($T \gg 1$) the population falls into a state of opinion bipolarization, where there are two groups of similar size with opposite opinions.  

The results presented in this article correspond to a fully connected network.  Although we expect that the conclusions remain valid qualitatively for other interaction topologies, it might be worthwhile to study the model in complex networks like scale-free or Erd\"os Renyi networks, which better represent social interactions.  It might also be interesting to explore how the stochastic resonance effect depends on the topology of the network.   

\section*{ACKNOWLEDGMENTS}

The authors are grateful to CONICET (Argentina) for continued support.

\end{document}